# Shadow-Oriented Tracking Method for Multi-Target Tracking in Video-SAR

Xiaochuan Ni, Xiaoling Zhang, Xu Zhan, Zhenyu Yang, Jun Shi, Shunjun Wei, Tianjiao Zeng
University of Electronic Science and Technology of China
Chengdu, China
Email: {hsiaochuan, zhanxu, zhenyuy@std.uestc.edu.cn}, {xlzhang, shijun, weishunjun, tzeng@uestc.edu.cn}

*Abstract*—This work focuses on multi-target tracking in Video synthetic aperture radar. Specifically, we refer to tracking based on targets' shadows. Current methods have limited accuracy as they fail to consider shadows' characteristics and surroundings fully. Shades are low-scattering and varied, resulting in missed tracking. Surroundings can cause interferences, resulting in false tracking. To solve these, we propose a shadow-oriented multi-target tracking method (SOTrack). To avoid false tracking, a pre-processing module is proposed to enhance shadows from surroundings, thus reducing their interferences. To avoid missed tracking, a detection method based on deep learning is designed to thoroughly learn shadows' features, thus increasing the accurate estimation. And further, a recall module is designed to recall missed shadows. We conduct experiments on measured data. Results demonstrate that, compared with other methods, SOTrack achieves much higher performance in tracking accuracy-18.4%. And ablation study confirms the effectiveness of the proposed modules.

## I. INTRODUCTION

Video synthetic aperture radar (Video-SAR) is capable of continuously monitoring observation scenes [1]. Thus, it plays a significant role in moving target tracking. Recent studies utilize shadows for moving target tracking [1]–[4]. Compared to the targets themselves, their shadows don't suffer defocusing and position offset problems. In another work, shadows focus on the actual locations, making it convenient for moving target tracking.

Generally, the most common processing framework contains two main steps: detection and association. The target is detected first, and then the results in different frames are associated together to generate a whole trajectory, known as track-by-detection (TBD) [5]. Therefore, the tracking accuracy is highly dependent on the detection performance. With the development of deep learning, shadow detection methods based on the deep network have achieved state-of-art performance. due to its strong ability to extract features from shadows. For example, a shadow detection method based on Faster-RCNN [6]and Bi-LSTM is proposed in [2]. And later, an improved LSTM-based [7] detection method is proposed in [3]. Taking another type of network, [1] proposed a detection-tracking combined end-to-end method based on SiamFc [8].

However, the above methods are all for single shadow tracking. For multi-target tracking, the problem is much more complex. Multiple different shadows are needed to be detected and associated. And situations like old shadows vanishing and new shadows emerging make it more complicated. Currently, researches related to multi-shadow tracking are few [9], [10]. And they are mainly based on multi-target tracking methods initially designed for optical videos. We think targets in optical videos and shadows in SAR videos have significant differences. Directly utilizing these methods would cause limited tracking accuracy. An example is shown in Fig. 1. Firstly, the shadows are small and faint; their shapes would change related to the location and velocity, thus causing missing detection and tracking. Secondly, both the shadows generated by stationary targets and the low-scattering area of clutter and noises may cause interferences, thus causing false detection and tracking.

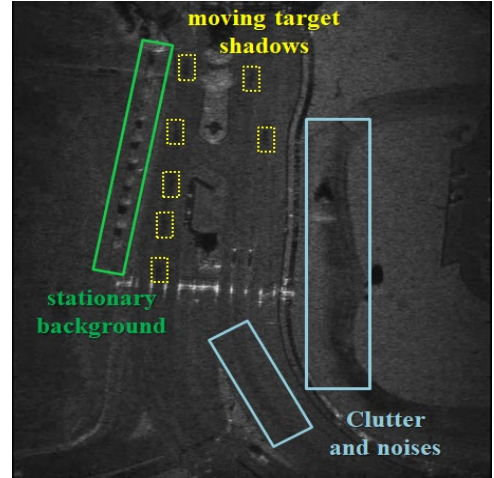

Fig. 1. Shadows in a complex environment.

Faced with these challenges and issues, we propose a shadow-oriented tracking method for multi-target tracking in Video-SAR, named SOTrack. The tracking method contains four phases: preprocessing, detection, association, and postprocessing. First, at the preprocessing phase, considering surroundings easily interfering with the shadows that cause false detection, we propose a multi-term spatial decomposition module that utilizes their features' differences to suppress the interferences and to enhance the shadows. Second, at the detection phase, considering the shadows are small that cause missing detection, sufficient feature extraction is the key. Thus, a state-of-art anchor-free detection network YOLOX [11] is introduced. And further considering their variable shapes, a multi-level spatial fusing module is proposed to enhance features. Third, at the association phase, considering the shadows' detection results may tend to be low-confidence which causes missing association, a low-confidence shadow recall module is proposed to reduce missing association. Lastly, at the postprocessing phase, a Gaussian interpolation module is performed to optimize the trajectory to increase the continuity and smoothness. An ablation study is conducted for the proposed modules to confirm their effectiveness.

Experiments on Sandia National Laboratories (SNL) video-SAR data [12] demonstrate that our method performs better than other state-of-the-art methods, in terms of tracking accuracy, including more accurate tracking and less false tracking. And the generated trajectories are more continuous and smoother. Compared with the state-of-art method, our method achieves a 18.4% accuracy gain.

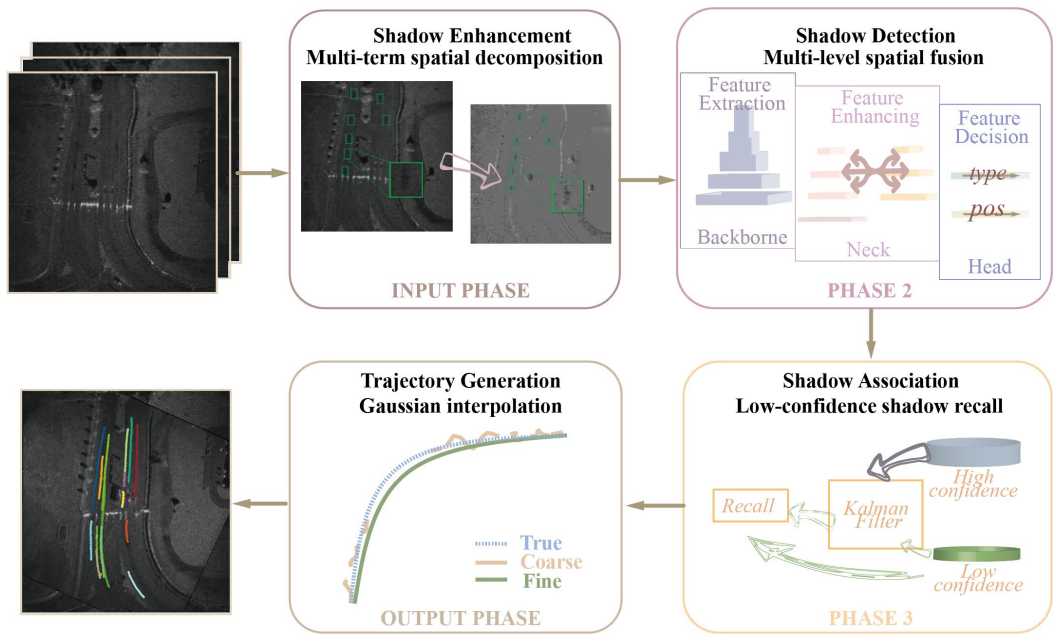

Fig. 2. The architecture of the proposed method.

## II. SOTRACK: SHADOW-ORIENTED TRACKING

The objective is to achieve more accurate multi-moving-target tracking in Video-SAR, where accurate and smooth trajectories are explicitly obtained. Adhering to it, we propose a shadow-oriented tracking method, where the shadow's characteristics are thoroughly considered to track more accurate and fewer false shadows in the entire tracking process. Following the typical tracking framework, it consists of two phases, shadow detection, and shadow association. Besides, shadow enhancement at the input and the trajectory generation at the output are also included. The architecture is shown in Fig. 2. Details are described in the following subsections.

### A. Shadown Enhancement with Spatial Decomposition

As surrounding clutters and noise are also low-scattering, and stationary background may form shadows, they all may cause false detection and tracking results. To address this, in the input phase, we propose to preprocess the Video-SAR data through multi-term spatial decomposition (MTSD), thoroughly utilizing the spatial-characteristic differences of the shadow term, the clutter-and-noise term, and the stationary background term. Specifically, the shadows are moving, largely variable, and sparse. The stationary background is relatively static, highly relevant, and low-rank. And the clutter-and-noise obeys the Gaussian distribution. Thus, the shadows can be enhanced by decomposition as follows.

$$\min_{\mathbf{L},\mathbf{X},\mathbf{N}} rank(\mathbf{L}) \ s.t. \ \mathbf{D} = \mathbf{L} + \mathbf{X} + \mathbf{N}$$
$$||\mathbf{X}||_0 \leq \delta, ||\mathbf{N}||_F \leq \sigma$$

where $\mathbf{D}$ is the input Video-SAR data, $\mathbf{L}$ is the background term, $\mathbf{X}$ is the shadow term, and $\mathbf{N}$ is the noise-clutter term. $rank()$ denotes the rank of input matrix, $|| \ ||_0$ denotes the $l_0$ norm, and $|| \ ||_F$ denotes the Frobenius norm. The problem can be solved with the proximal gradient algorithm [13]. Details can be found in our previous work [14]. After multi-term spatial decomposition, the contrast between shadows and the background is enhanced a lot, and the interference from the noise and clutter and the static shadow is also reduced. An example is given in Fig. 3.

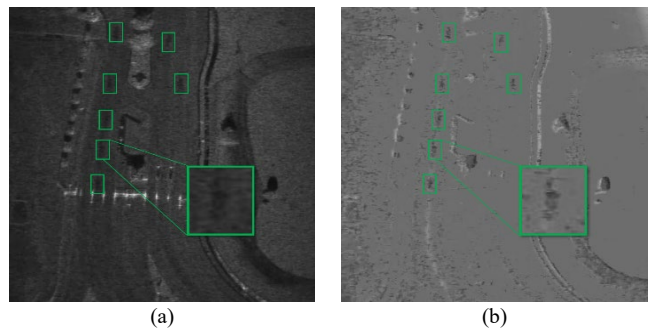

Fig. 3. An example of the preprocessing of shadow enhancement at a single frame. Different moving target shadows are marked in green boxes. (a) Before enhancing. (b) After enhancing.

### B. Shadow Detection with Spatial Fusion

The detection phase lays the foundation for tracking. The shadow position information is obtained through detection in every frame of enhanced Video-SAR data. Then it is fed into the next association phase to generate coarse trajectories. Thus, the final precision of tracking and the quality of the generated trajectories are highly dependent on this phase.

As the shadows are generally small, thus there may be lots of missing detection. To address this, the deep-learning-based detection method is adopted. Among them, methods with the anchor mechanism generally have high detection precision. Anchors, taking the role of guidance, are dense bounding boxes to locate the potential locations of targets [15].

Such a mechanism is suitable for the scenario where dense targets exist. However, as seen in Fig. 3, shadows are highly sparse. If we adopt the anchor mechanism, most anchors would be invalid and may even cause false detections.

In this sense, considering the sparse characteristic of shadows, we adopt the anchor-free detection method, where the recently proposed YOLO-X [11] detection architecture is introduced. It mainly relies on the differences between targets and non-targets to detect. So, it consists of feature attraction (backbone), feature enhancement (neck), and feature decision (head). Two aspects are featured. First, in the feature enhancement phase, multilevel spatial features are learned to compensate for the spatial information loss without anchors. Second, in the feature decision phase, decoupled detection head is adopted to balance the dual tasks of position learning and target/non-target type learning, avoiding the propagation cumulative errors from coupled head.

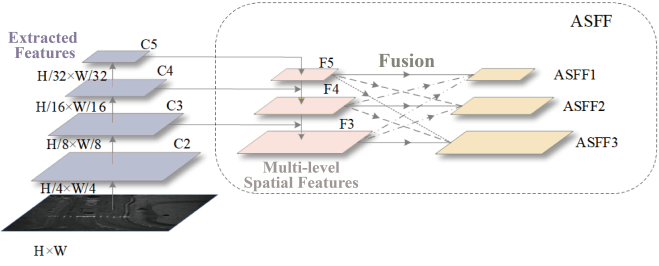

Fig. 4. Adaptively Spatial Feature Fusion (ASFF) module.

However, it is originally proposed for optical video, directly applying this architecture is not appropriate. The scale in Video-SAR tends to change due to the velocity and position. This corresponds to the multi-scale detection ability for detection. The original feature enhancement module in YOLO-X enhances this ability with multi-level spatial feature learning, which may not be suitable for relatively fast scale variation of shadows in Video-SAR. This would result in the missing detection of shadows.

To address this issue, in addition to the multi-level spatial feature learning, we improve it into multi-level spatial fusion, where the multi-level features are comprehensively fused. The improved feature enhancement module is dubbed Adaptively Spatial Feature Fusion (ASFF). And its architecture is shown in Fig. 4. The extracted multi-level spatial features are fed into ASFF, where two feature flows are designed to enhance the features. The first flow is from top to bottom, gradually decreasing the receptive field. During this flow, the attention is also gradually focused from the shadow's self-features to the surrounding features. The second flow is adaptively multi-level spatial features fusion, where different level features are mixed with each other, which reduces the less effective features at the self-level with the help of other levels. The original feature maps of F3, F4, and F5 are fused into ASFF1, ASFF2, and ASFF3. This fusion can be expressed as.

$$X_{i,j}^{n \to l}$$

Where $X_{i,j}^{n \to l}$ denotes the feature at the position $(i,j)$ of the feature map shaped from the $n$th-level map to the $l$th-level map. The shaping is conducted through up/down sampling to match the sizes. $Y_{i,j}^l$ denotes the fused feature at the corresponding position. $\alpha_{i,j}^l, \beta_{i,j}^l$, and $\gamma_{i,j}^l$ are the fusing weights of different levels. They are obtained through $1 \times 1$ convolution and the softmax operation on the reshaped map. These weights are learned during the network training phase.

### C. Shadow Association with Low-confidence Recall

The association phase is the critical linking part for tracking. The detection output and coarse trajectories are linked through it. The discrete detected bounding boxes are associated into separate continuous trajectories. This is normally achieved by Kalman Filter [16]. However, directly applied Kalman Filter as usual, low-confidence detection results would be filtered out. These detection results exist quite a few.

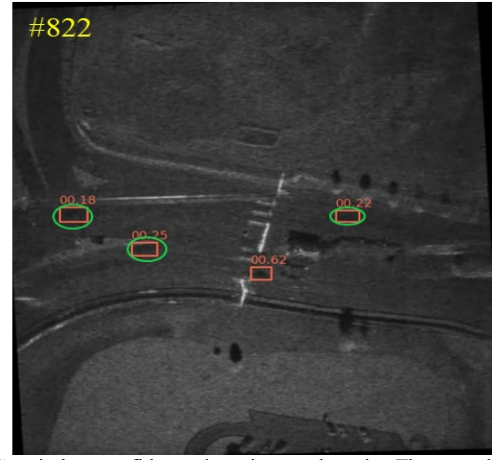

Fig. 5. Certain low-confidence detection results exist. The green bounding boxes are the ground truth of shadows. The orange ones are the detection results with confidences.

An example is illustrated in Fig. 5. The causes of low-confidence are mainly due to the occlusion, blurring or glinting of shadows in different frames. These filtered out results can cause missing associations, divided trajectories, and may even wrong trajectories.

To avoid these missing associations and form more continuous trajectories, in addition to the original Kalman Filter, inspired by recent work ByteTrack [16], we improve it with additional two-phases low-confidence shadow recall. The recall mechanism in ByteTrack only recall the low-confidence in the output phase of Kalman Filter. We modified it into two phases, considering low-confidence shadow both in the input and output phases, as shown in Fig. 2. Specifically, in the input phase, both low-confidence detection results and high-confidence detection results are fed into Kalman Filter. It associates the current frame's shadow position through estimation according to the previous frames' result, and further adjust according to the confidence of the current frame's shadow detection result. For example, at the frame $k$, the association process can be expressed as.

$$\begin{aligned}
\overline{\mathbf{x}}_k &= \mathbf{A}\mathbf{x}_{k-1} \\
\overline{\mathbf{P}}_k &= \mathbf{A}\mathbf{P}_{k-1}\mathbf{A}^\mathrm{T}\mathbf{x}_{k-1} + \mathbf{Q} \\
\mathbf{y}_k &= \mathbf{z}_k - \mathbf{H}\overline{\mathbf{x}}_k \\
\mathbf{K}_k &= \overline{\mathbf{P}}_k \mathbf{H}^\mathrm{T}[\mathbf{H}\overline{\mathbf{P}}_k \mathbf{H}^\mathrm{T} + (1-c_k)\mathbf{R}]^{-1} \\
\mathbf{x}_k &= \overline{\mathbf{x}}_k + \mathbf{K}_k \mathbf{y}_k
\end{aligned}$$

This weight $c_k$ is influenced by the confidence. If the detection confidence is high, meaning the uncertainty is low, and the detection noise is small, then the weight is large adaptively. Then, in the output phase, the low-confidence detection results are recalled again. Specially, first, for those detection results with the high confidence, higher than a predefined threshold, they are measured with the results of Kalman filter by the similarity metric Intersection Over Union (IOU). If the measuring results are higher than 0.5, we consider these detection results as new parts of trajectories. For those that are less than 0.5, but exist in several adjacent frames, we consider these detection results as new trajectories. Second, for those detection results with low confidence, they are measured with the left-out results of Kalman filter at the last step. Still, if the results are higher than 0.5, we recall them as parts of trajectories. All the other detection results and Kalman Filter's results that are unmatched are dropped. At last, we illustrate the whole association phase in Fig. 6.

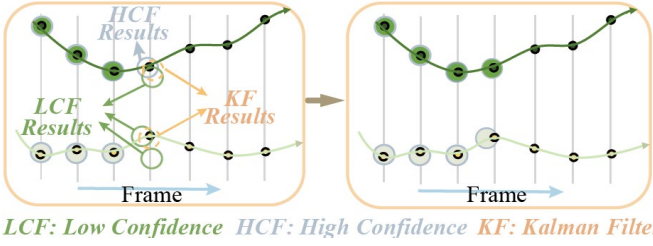

Fig. 6. Association with Low-confidence Shadow Recall.

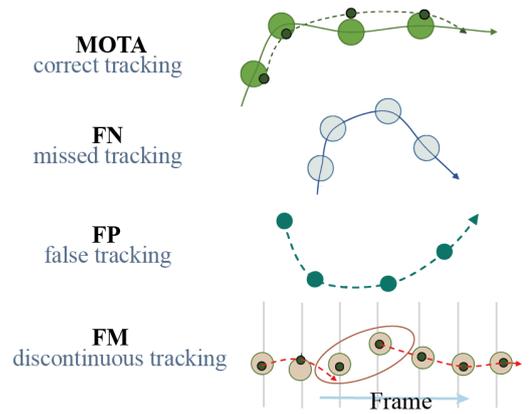

Fig. 8. Metrics adopted for different tracking scenarios.

TABLE I. ABLATION STUDY RESULTS

| Mt SD | Ml SF | Lc SC | GI | MOTA | FP | FN | FM |
|---|---|---|---|---|---|---|---|
| - | - | - | - | 36.96 | 283 | 779 | 106 |
| ✓ | ✓ | - | - | 53.82% (↑16.9%) | 160 | 616 | 96 |
| ✓ | ✓ | ✓ | - | 54.82% (↑17.9%) | 150 | 608 | 96 |
| ✓ | ✓ | ✓ | ✓ | 55.35% (↑18.4%) | 175 | 574 | 90 |

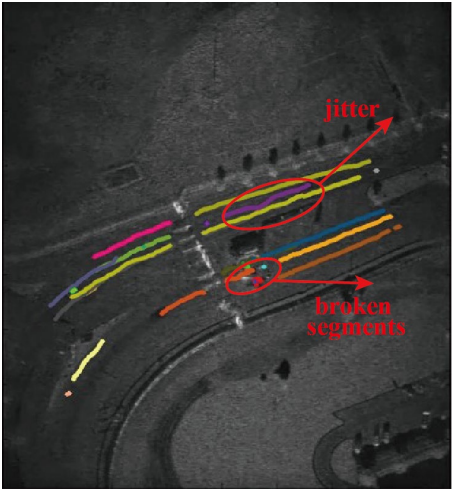

Fig. 7. Coarse trajectories after the association phase. Trajectories are jittered and broken into segments.

### D. Trajectory Generation with Gaussian Interpolation

The trajectories from the last phase are coarse in term of smoothness and continuous. Although in the previous phases, efforts are spared to detect and associate more accurate targets, there are still residual noises causing detected shadow positions' deviation and jitter. And residual stationaries exist in the way of moving shadows causing occultations and broken segments of trajectories. An example is given in Fig.7.

To address these issues, we adopt additional processing at the output phase by Gaussian interpolation. Compared to the simple linear interpolation, it suits more faced with the scenario that the trajectory is not straight but curved. The details of standard Gaussian interpolation can be found in [17].

### III. EXPERIMENTS

We validate the proposed method in this section. Experiments include ablation studies on the proposed modules. And comparison with other tracking methods is conducted. Details are as follows.

### A. Dataset and Metrics

All the experiments are conducted on the Sandia National Lab Video-SAR dataset. It contains 899 frames that form 9 videos. 6 of them are used for training, and the remained 3 ones are used for testing. The objective of multi-target tracking is to obtain all the targets' trajectories that are accurate, continuous and smooth. Thus, 5 metrics are adopted, including (Multiple Objects Tracking Accuracy), FP (False Positives), FN (False Negatives), and FM (Fragmentation) [18]. The first one is related to the tracking accuracy, thus the higher is better. The others ones are related to the false tracking, missed tracking, and discontinuous trac- king. Thus, the lower is better. Fig.8 illustrates the scenarios related to these metrics.

### B. Ablation Studies

Compared with the methods that's are intended for optical images, in this work, four additional improvements are proposed to multi-target tracking in Video-SAR with the shadow-oriented principle. To verify their effectiveness, we conduct ablation studies. The baseline method adopts the recent proposed ByteTrack that achieve the state-of-art performance for optical videos. The four improvements in the corresponding four phases are abbreviated as MtSD (Multi-term Spatial Decomposition), MlSF (Multi-level Spatial Fusion), LcSC (Low-confidence Shadow Recall), and GI (Gaussian Interpolation). The results are listed in Table 1. The ablation study on the MtSD only has been fully conducted in our previous work. For the sake of simplicity, only results of other three improvements are shown.

It can be seen that MOTA is increased by 16.9% with the MtSD and the MlSF, and FP and FN are decreased by 123 and 163, respectively, indicating that false tracking and missed tracking are reduced a lot after shadows are enhanced and features are extracted more efficiently. At the same time, FM is decreased by 10, indicating that the trajectory interruption happens much less due to the reduction of missed tracking.

Then with the LcSC, MOTA is further increased by 1.00%, and the FP and FN are decreased by 10 and 8 respectively. As the Kalman filter is adaptively adjusted by the detection confidence, the shadows' locations are more accurate.

At last, after Gaussian interpolation, MOTA is further increased by 0.5%, and FN and FM are decreased by 34 and 6, respectively. It shows that the trajectory GI can reasonably fill the gaps of trajectory, thus reducing the number of missed tracking and broken trajectory segmentations. In all, the proposed four shadow-oriented improvements achieve 18.4% MOTA increasement, indicating the tracking precision is increased a lot.

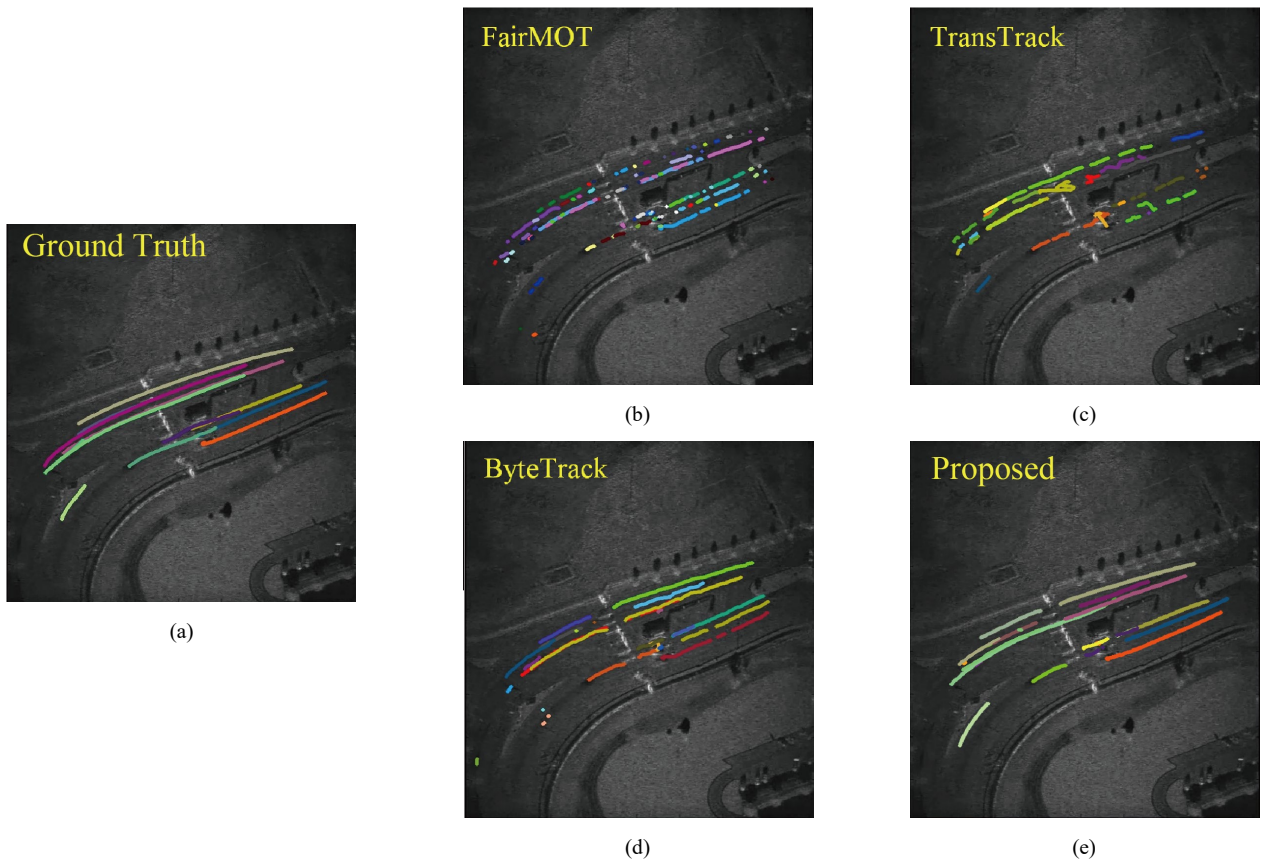

Fig. 9. Visualization results of different method. (a) Ground truth. (b) Result of FairMot. (c) Result of TransTrack. (d) Result of ByteTrack. (e) Result of the proposed method, SOTrack.

TABLE II. COMPARISON WITH OTHER METHODS

| Method | MOTA | FP | FN | FM |
|---|---|---|---|---|
| TransTrack | 31.26 | 242 | 905 | 138 |
| FairMOT | 34.67 | 223 | 786 | 160 |
| ByteTrack | 36.96 | 283 | 779 | 106 |
| SOTrack (proposed) | 55.35% (↑18.4%) | 175 (↓21.5%) | 574 (↓26.3%) | 90 (↓15.1%) |

### C. Comparisons with Other Methods

Three state-of-the-arts tracking methods are compared with. They are TransTrack, FairMOT, and ByteTrack. The results are shown in Table 2, which shows that the proposed method achieve the highest performance in all aspects, with an averaged 20.3% performance incensement. With the shadow-oriented principle, we take the shadow characteristics into consideration through the whole processing phases, and result the multi-target tracking performance can be increased a lot.

### D. Visualization Results

At last we present some visualization results of difference methods in Fig. 9. Among them, FairMOT has the worst performance in terms of trajectories' continuity. Nearly all the trajectories are broken into short segments. TransTrack has a little better performance. However, the tracking accuracy is worse that some of the trajectories are twisted out of shapes. Compared with the former two, the performance of ByteTrack is much better. Much continuous trajectories are obtained, but they are jittered in general. And some false tracking results exist in the left-bottom corner. Our method obtains the best performance. Less false tracking results exist, and more accurate, smooth, and continuous trajectories are obtained. In general, the visualization results match the quantitative results in the tables.

## IV. CONCLUSION

A shadow-oriented tracking method, SOTrack, for multi-target tracking in Video-SAR. As current methods are based on methods initially designed for optical videos, causing limited tracking accuracy, we consider SAR video characteristics more thoroughly. Specifically, we develop the tracking method with an orientation toward shadows and their surroundings. Faced with missing and false tracking issues resulting from them, under the classical track-by-detection framework, four designed modules are adopted to address them. Among them, the first module in the prepossessing is to suppress the surroundings' interferences and enhance shadows, which utilize their spatial features' differences. The latter two react by reducing missing tracking, where more accurate detection and association are needed. Thus, an improved deep-learning-based detection network for small-scale and variable shadows is designed. And a shadow recall module is intended for the association, where low-confidence detected shadows are recalled. At last, for more continuous and smoothers trajectories, Gaussian interpolation is introduced. The ablation study confirms their effectiveness, especially for the first enhancing module. Compared with other methods, the tracking performance increases significantly, with more accurate tracking, less false tracking, and more intact trajectories.